\documentclass[lettersize,journal]{IEEEtran}
\usepackage{graphicx} 
\usepackage[export]{adjustbox} 
\usepackage{xcolor}
\usepackage{enumerate}
\usepackage{amssymb}
\usepackage{float}

\usepackage[normalem]{ulem}
\usepackage{epstopdf}

\usepackage{bbm}
\usepackage{cite}
\usepackage{amsmath}
\newtheorem{theorem}{Theorem}

\newtheorem{proposition}{Proposition}

\newtheorem{definition}{Definition}
\newtheorem{lemma}{Lemma}
\newtheorem{remark}{Remark}
\title{Coalitional Game Framework for Multicast in Wireless Networks}

\author{Anjali Yadav, Arya Agarwal, Alok Kumar,  Tushar S. Muratkar, Gaurav S. Kasbekar
\thanks{A. Yadav, A. Agarwal, A. Kumar, and G.S. Kasbekar are with the Department of Electrical Engineering, Indian Institute of Technology Bombay, Mumbai 400076, India. T. S. Muratkar is with the Department of Electronics \& Communication Engineering at Indian Institute of Information Technology, Nagpur 441108, India. Their e-mail addresses are anjaliyadavay3008@gmail.com, agarwalarya29@gmail.com, vibhutimayank@gmail.com, gskasbekar@ee.iitb.ac.in, and tushar\_16m@yahoo.co.in.}}

\begin{document}
\maketitle
\begin{abstract}
 We consider a wireless network in which there is a transmitter and a set of users, all of whom want to download a popular file from the transmitter. Using the framework of cooperative game theory, we investigate conditions under which users have incentives to cooperate among themselves to form coalitions for the purpose of receiving the file via multicast from the transmitter.  First, using the solution concept of core, we investigate conditions under which it is beneficial for all users to cooperate, i.e., the grand coalition is stable. We provide several sets of sufficient conditions under which the core is non-empty as well as those under which the core is empty. Next, we use the concept of $\mathbb{D}_c$-stability to identify a set of sufficient conditions under which the users in the network form a certain fixed number of coalitions such that all the users within each coalition cooperate among themselves. Our analytical results show how the values of different system parameters, e.g., data rates of different users, transmit and receive power, file size, bandwidth cost, etc., influence stability properties of coalitions, and provide a systematic approach to evaluating cooperation of users for multicast. We also study cooperation among different users using numerical computations. The problem of coalition formation in the context of multicast addressed in this paper is fundamental, and our analysis provides new insights into the feasibility of stable cooperative multicast strategies, contributing to a deeper understanding of cooperation in wireless networks.
\end{abstract}


\section{Introduction}
As the demand for high-speed wireless data services continues to surge, the efficient distribution of content, e.g., popular videos and other files, has become increasingly critical \cite{xing2022secure}. Unicast methods, which involve separately transmitting content to each user, are inefficient and unsustainable when a given file needs to be distributed to a large number of users in a wireless network. In such scenarios, \emph{multicast} techniques, which enable simultaneous data transmission to multiple users, can enhance spectral efficiency, reduce latency, optimize the use of available bandwidth, reduce network congestion, and help in managing network resources more effectively \cite{varshney2002multicast}.

Consider a wireless network in which a transmitter needs to distribute some popular content to multiple selfish and rational users. For example, a base station (BS) (respectively, satellite) may need to distribute a popular file to cellular users (respectively, ground terminals) in a wireless cellular network (respectively, satellite network) \cite{ma2023resource}. When the transmitter multicasts the file to a subset of the users, it is often beneficial for the users to cooperate and share the costs incurred in the multicast operation (e.g., energy consumed in transmitting and receiving the file, bandwidth used) among themselves. Such a set of users who cooperate among themselves is called a \emph{coalition} \cite{r1}. By forming coalitions, users can collaborate to optimize their utilities, use resources more effectively, and enhance the overall efficiency of content distribution. \emph{Coalitional game theory} \cite{r1} enables us to analyze this cooperation, providing insight into how users can dynamically form coalitions to maximize their utilities, while minimizing costs \cite{saad2009coalitional,maiti2024abp}.

Several prior works have studied the optimization of multicast through advanced scheduling and beamforming techniques \cite{varshney2002multicast, li2024multicast,won2009multicast,low2010optimized,vella2012survey, li2019energy,li2020low,du2021capacity}, and the application of coalitional game theory for resource allocation and cooperative content delivery in wireless networks \cite{saad2009coalitional,maiti2024abp,li2014coalitional,hou2020overlapping,zhang2018context,chen2018cvcg,r3_new,r6_new,r7_new,r8_new,wang2013dynamic,r7,alabiad2020coalition,militano2015constrained,zhou2018dependable,wang2020coalition,liu2025hedonic,chou2024gaussian} (see Section \ref{SC:related:work}). However, in prior work,   the conditions under which coalitions formed for multicast in wireless networks are stable  have not been analyzed. Most existing research focuses on scheduling, resource allocation, and achieving throughput enhancements, without investigating the emergence of stable cooperative systems for multicast.  In particular, little research has been done on the circumstances in which the \emph{grand coalition} (set of all users in the network) either stays stable or splits up into smaller, stable coalitions.
We address this gap in this paper by applying solution concepts from coalitional game theory such as \emph{core} \cite{r1} and \emph{$\mathbb{D}_c$-stability} \cite{apt2009generic} to the problem of multicast in wireless networks, providing new insights into stable coalition formation in this context. To the best of our knowledge, this paper is the first  to use cooperative game theory to systematically analyze the stability of coalitions formed for cooperative multicast in wireless networks by deriving explicit analytical conditions under which the grand coalition either stays stable or breaks up into smaller, stable coalitions.

In this paper, we consider a wireless network in which there is a transmitter and a set of users, all of whom want to download a popular file from the transmitter. 
Our system model is general and applicable to a variety of wireless networks, e.g.: (i) a BS in a cellular network may need to distribute a file to its associated users; (ii) a satellite in a satellite network may distribute a file to a set of ground terminals; (iii) a gateway in an Internet of Things (IoT) network may distribute a file to IoT devices; (iv) a road side unit (RSU) in a vehicular network may distribute a file to vehicles; and (v) a ground station in a drone network may distribute a file to the drones in its range. We investigate conditions under which users have incentives to cooperate among themselves to form coalitions for the purpose of receiving the file via multicast from the transmitter. The data rates at which different users can download data from the transmitter are, in general, different. The \emph{value} \cite{r1} of a coalition is the sum of the valuations that the users derive from obtaining the file minus the energy and bandwidth costs incurred during the multicast operation. We model the above problem using the framework of cooperative game theory \cite{r1}. Our analytical results are divided into two parts:  
\begin{itemize}
    \item 
    In the first part, using the solution concept of core \cite{r1} from cooperative game theory, we investigate conditions under which it is beneficial for \emph{all} the users to cooperate, i.e., the grand coalition is stable. We consider two important special cases of this game: 
    \begin{itemize}
        \item 
        In the first special case, we consider a network where the data rates of all the users are equal. We show that the core is always non-empty (and hence the grand coalition can be stabilized).
        \item 
        In the second special case, we consider a generalization in which the data rates at which different users can download data from the transmitter, as well as the power consumed while doing so, lie in ranges with certain upper and lower limits. We show that the core is non-empty when the upper and lower limits of the above ranges satisfy certain conditions.
    \end{itemize}
    Next, we identify two special cases in which the core is empty under certain conditions: 
    \begin{itemize}
        \item 
        In the first special case, we show that when the ratio between the second-minimum and minimum data rate of any user in the network exceeds a certain value, the core is empty (and hence the grand coalition cannot be stabilized).
        \item 
        Similarly, in the second special case, we show that when the ratio of the maximum and minimum data rate of any user exceeds a certain value, the core is empty.
    \end{itemize}
    \item 
 In the second part, we use  the concept of $\mathbb{D}_c$-stability \cite{apt2009generic} to identify a set of sufficient conditions, in terms of the data rates, transmit power, receive power of different users, and bandwidth cost, under which the users in the network form a certain fixed collection of coalitions such that all users within each coalition cooperate among themselves. 
\end{itemize}
Our analytical results show how the values of different system parameters influence stability properties of coalitions, and provide a systematic approach to evaluating cooperation of users for multicast. We also study cooperation among different users using numerical computations. Our numerical results provide several insights. E.g., they show that  coalition preferences vary significantly with the network parameters: separate downloads are preferred when there is a high variation among the rates of different users, $D_c$-stability is observed in the regime of medium variation of the rates, and the grand coalition becomes favorable when the rates are close to each other. The problem of coalition formation in the context of multicast addressed in this paper is fundamental, and our analysis provides new insights into the feasibility of stable cooperative multicast strategies, contributing to a deeper understanding of cooperation in wireless networks.

The rest of the paper is organized as follows. We provide a review of related work in Section \ref{SC:related:work}. In Section \ref{SC:background}, we present some background concepts.  
In Section \ref{SC:system:model:problem:formulation}, we describe our system model and problem formulation. In Section \ref{SC:core:non:empty} (respectively, Section \ref{SC:core:empty}), we prove that under certain sets of conditions, the core is non-empty (respectively, empty). In Section \ref{SC:Dc:stability}, we prove that under certain conditions, a set of coalitions is $\mathbb{D}_c$-stable. We provide numerical results in Section \ref{SC:numerical:results} and conclusions and directions for future research in Section \ref{SC:conclusions:future:work}.

\section{Related Work}
\label{SC:related:work}
In prior work, researchers have proposed different strategies for multicast in both wired and wireless networks; these strategies have been designed for achieving various goals, e.g., to enhance spectral efficiency and to reduce network congestion. In \cite{varshney2002multicast}, the fundamental challenges involved in multicasting over wireless networks were identified and issues such as channel fading and dynamic user mobility were highlighted. In \cite{vella2012survey}, a comprehensive overview of multicast protocols, with an emphasis on the trade-offs between efficiency, reliability, and overhead, was provided. In \cite{li2024multicast} and \cite{won2009multicast}, advanced scheduling techniques, including those based on deep reinforcement learning, were leveraged to optimize multicast performance. In \cite{low2010optimized}, opportunistic multicast scheduling strategies to maximize the throughput while minimizing the resource consumption were proposed. In \cite{li2020low}, multicast beamforming techniques for millimeter wave (mmWave) communications were proposed, and significant improvements in energy efficiency and network capacity were demonstrated. In \cite{li2019energy}, an energy-efficient precoding design for a multi-user multiple-input multiple-output (MIMO) downlink system, which uses Layered-Division Multiplexing (LDM) to support hybrid multicast and unicast services, was investigated. The authors formulated an energy efficiency (EE) maximization problem under both multicast and unicast data rate constraints, which is challenging due to its non-smooth and non-convex nature. To address it, they proposed a low-complexity first-order algorithm that relies only on gradient information. 
In \cite{du2021capacity},  reconfigurable intelligent surfaces (RIS) were used to enhance multicast transmission in a system where a multi-antenna BS serves multiple single-antenna users. An equivalent channel model was developed and an optimization problem to maximize capacity through joint design of the transmit covariance matrix and RIS phase shifts was formulated. A gradient descent-based alternating optimization approach was proposed for general scenarios, while a globally optimal solution was derived for a special case. The study also examined how capacity scales with increasing numbers of RIS elements, BS antennas, and users. In \cite{10930664}, a greedy clustering heuristic and a binary integer programming framework were proposed for multicast resource block allocation in beyond 5G (B5G) networks, which improve the quality of service (QoS) and spectral efficiency over baseline techniques.

Coalitional game theory has been extensively used for resource allocation, spectrum sharing, and cooperative communication in wireless networks. In \cite{saad2009coalitional}, an extensive review of applications of coalitional game theory to communication networks was provided. In \cite{li2014coalitional}, coalitional game theory  was applied for resource allocation in device-to-device (D2D) networks, and improved spectral efficiency was demonstrated. In \cite{zhang2018context},  a coalition formation framework was proposed for context-aware group buying in D2D networks, while in \cite{hou2020overlapping}, a multicast scheme using overlapping coalitions to optimize data delivery in backhaul-limited networks was proposed. In \cite{chen2018cvcg}, cooperative vehicle-to-vehicle (V2V) communication was studied using a coalitional game approach. In \cite{xing2022secure}, coalitional game theory was used to secure the delivery of content to connected and autonomous vehicles. In \cite{r3_new}, the problem of enhancing cell-edge user service in 5G small-cell networks was studied by examining coalition building games. In contrast to static and greedy clustering algorithms, the study improves edge user throughput by addressing interference issues. In \cite{r6_new}, joint optimization of resource allocation and mobile device association in small cell  IoT networks with edge computing and caching was studied.  The proposed technique reduces weighted delays and improves resource usage using convex optimization and a coalitional game. In \cite{r7_new}, resource allocation was performed in heterogeneous D2D communication networks with RIS assistance to optimize the overall rate of the system. The problem was decomposed into subproblems for phase conversion, power allocation, and resource allocation, and these subproblems were addressed using local search methods, gradient descent, and coalitional game theory, respectively. In \cite{r8_new}, a distributed cooperative spectrum sensing (CSS) method based on a greedy coalitional game was proposed for cognitive radio networks. A heuristic, which strikes a compromise between throughput, cost, and energy overhead, was proposed for solving the NP-hard problem formulated in the paper. In \cite{10461128},  the RIS phase-shift configuration and multicast user grouping problem was formulated as a  coalition construction game. A utility function based on the signal-to-interference-plus-noise ratio (SINR) and energy efficiency was used, and a distributed method, which utilizes merge-and-split criteria for the building of stable coalitions, was proposed.

In \cite{wang2013dynamic}, a coalition-based cooperative approach for content distribution in vehicular ad hoc networks (VANETs), where vehicles download file pieces from a RSU and share them with neighbors, was proposed. The method improves the performance of the content delivery process despite high mobility and poor vehicle-to-RSU (V2R) channels. In \cite{alabiad2020coalition,militano2015constrained,zhou2018dependable}, efficient content distribution and uploading in D2D networks using cooperative strategies and coalitional game theory was explored. In \cite{alabiad2020coalition}, the focus is on delay minimization in decentralized D2D networks through network coding and a distributed coalition formation algorithm.  In \cite{militano2015constrained},  content distribution in vehicular D2D networks is enhanced by leveraging trajectory prediction to form stable, delay-minimizing content-sharing groups. In \cite{zhou2018dependable},  a multihop D2D relay-based uploading scheme, where user devices form chains to forward content to the evolved NodeB (eNodeB), was investigated. This chain formation is modeled as a constrained coalitional game, reducing the upload delays compared to conventional cellular methods. A semantic-aware coalition formation game for D2D multicast was proposed in \cite{10840352}. To avoid duplication and boost the spectral efficiency, users with comparable content requirements band together. It was shown that this method results in improvements in the throughput, energy efficiency, and semantic satisfaction. Together, these works \cite{alabiad2020coalition,militano2015constrained,zhou2018dependable,10840352}  highlight the effectiveness of coalition-based cooperation in improving the delivery performance and network efficiency in various D2D scenarios. In \cite{10949678}, a coalition formation game for ensuring secure communication and resistance to jamming in air-terrestrial ad-hoc networks was studied.  Stable coalitions are established between UAVs and users through the application of Pareto dominance and merge-and-split procedures.  The results in \cite{10949678} indicate that compared to non-cooperative approaches, secrecy and robustness are improved under the proposed cooperative approach.

In \cite{wang2020coalition}, the problem of popular content distribution in mmWave vehicular networks, with the objective of minimizing the communication burden on the BS, was addressed by leveraging full-duplex (FD) V2V communication. A scheme based on a coalition formation game  was proposed to maximize the number of files received by on-board units (OBUs), while ensuring fairness through transferable individual profits.  As the IoT and deep learning advance, UAV-based drones-as-a-service (DaaS) has emerged for distributed AI training. However, privacy concerns hinder collaboration among independent UAVs. To address this, in \cite{liu2025hedonic}, a federated learning framework based on hedonic coalition formation and contract theory was proposed. It enables secure and incentive-aligned cooperation among UAVs and IoT nodes. In \cite{chou2024gaussian}, the authors explored whether selfish transmitters in a Gaussian multiple access wiretap channel  can benefit from cooperation. Using coalitional game theory, the study showed that in degraded channels, cooperation is always advantageous and leads to a fair and stable secrecy rate allocation. In non-degraded settings, the benefits of cooperation depend on the channel parameters. In \cite{maiti2024abp},  a secure multicast architecture that combines coalitional game theory with attribute-based proxy re-encryption was proposed, which guarantees effective and incentive-compatible content delivery.  It supports equitable interaction among users using Shapley values. 

Despite the progress made in the above papers on multicast \cite{varshney2002multicast, li2024multicast,won2009multicast,low2010optimized,vella2012survey, li2019energy,li2020low,du2021capacity,10930664} and applications of coalitional game theory to wireless networks \cite{xing2022secure,saad2009coalitional,maiti2024abp,li2014coalitional,hou2020overlapping,zhang2018context,chen2018cvcg,r3_new,r6_new,r7_new,r8_new,wang2013dynamic,r7,alabiad2020coalition,militano2015constrained,zhou2018dependable,wang2020coalition,liu2025hedonic,chou2024gaussian,10461128,10840352,10949678,wang2020coalition}, none of these papers analyze the conditions under which coalitions formed for multicast in wireless networks are stable. Specifically, they do not investigate the circumstances in which the grand coalition either stays stable or splits up into smaller, stable coalitions.
In contrast, in this paper, we apply solution concepts from coalitional game theory such as core \cite{r1} and $\mathbb{D}_c$-stability \cite{apt2009generic} to the problem of multicast in wireless networks, and derive explicit analytical conditions under which the grand coalition either stays stable or breaks up into smaller, stable coalitions. 

The closest to this paper is our prior work \cite{r7}, in which a coalitional game framework was proposed for content distribution using D2D communication. The conditions under which it is beneficial for cellular users to cooperate among themselves were analyzed using the concepts of core and $\mathbb{D}_c$-stability. However, the system model and results of this paper are significantly different from those in \cite{r7}. First, in the system model in \cite{r7}, content is distributed from the BS in a cellular network to its associated users via relays, which communicate with the destination nodes using D2D communication. In contrast, in the system model in this paper, a transmitter directly multicasts a popular file to a set of users, and relays and D2D communication are not used. Second, the system model and results in \cite{r7} are applicable only to the specific scenario in which a BS in a cellular network that supports D2D communication distributes content to its associated users using relays. In contrast, the system model and results in this paper are much more general and applicable to any scenario in which a transmitter multicasts information to a set of users; e.g., in addition to the scenario of multicast from a BS in a cellular network to its associated users, they are also applicable to a satellite network in which a satellite multicasts information to a set of ground terminals, an IoT network in which a gateway multicasts information to IoT devices, a vehicular network in which a RSU multicasts information to vehicles, a drone network in which a ground station multicasts information to the drones in its range, etc. Third, since the system model in this paper is significantly different from that in \cite{r7}, the specific analytical results obtained for the two models and the techniques used to obtain them are also different. Fourth, in this paper, we prove several key results, which have no counterparts in \cite{r7}. For example, 
in \cite{r7}, a proof of non-emptiness of the core was provided only for a special case in which all D2D and BS-cellular user communication links are symmetric. In contrast, in this paper, apart from a symmetric scenario, we also show the non-emptiness of the core in a more general, asymmetric scenario (see Theorem \ref{TH:Rmax:by:Rmin:ub}). Also, in \cite{r7}, a proof of emptiness of the core was provided only for a small example network with six users. In contrast, in this paper, we derive multiple sets of sufficient conditions under which the core is empty in networks with arbitrary numbers of users (see Theorems \ref{TH:min:second:min:core:empty} and \ref{TH:min:max:core:empty}). In summary, the system model and results in this paper considerably differ from those in \cite{r7}. The analysis in this paper provides new insights into the feasibility of stable cooperative multicast strategies, contributing to a broader understanding of cooperative wireless networks.

\section{Background}
\label{SC:background}
\subsection{Coalitional Game}
Consider a system containing a set of players $\mathcal{N} = \{1,\ldots,N\}$.
A \emph{coalition} refers to a subset of players \( S \subseteq \mathcal{N} \) who cooperate among themselves \cite{r1}. The set of all players \( \mathcal{N} \) is referred to as the \emph{grand coalition}.
A \emph{transferable utility} coalitional game is defined by the pair $(\mathcal{N}, v)$, where $v(\cdot)$ is a function that assigns a real value to each coalition $S \subseteq \mathcal{N}$ \cite{r1}. The function $v(S)$ is known as the \emph{value function}, and equals the total utility or payoff that the players of coalition $S$ can collectively achieve when they cooperate among themselves \cite{r1}.

\subsection{Core}
In cooperative game theory, the \textit{core} \cite{r1} is a fundamental concept that helps us determine whether a coalition of players is stable. The core is the set of possible distributions of total utility among the players such that no subset of players has an incentive to break away and form their own coalition \cite{r1}.  

\begin{definition}
\label{DF:core}
The core is the set of all payoff profiles \( (x_1,  \ldots, x_N) \), where $x_i$ denotes the payoff of player $i$, which satisfy the following two conditions:
\begin{enumerate}[(1)]
    \item \textit{Efficiency:} The entire value of the grand coalition is distributed among the players:
    \begin{equation}
    \label{EQ:core:efficiency}
        \sum_{i \in \mathcal{N}} x_i = v(\mathcal{N}).
    \end{equation}
    
    \item \textit{Coalitional Rationality:} No subset of players has an incentive to deviate and form their own coalition:
    \begin{equation}
    \label{EQ:core:coalitional:rationality}
        \sum_{i \in S} x_i \geq v(S), \quad \forall S \subseteq \mathcal{N}.
    \end{equation}
    This condition ensures that each coalition $S$ receives at least as much utility as they could achieve on their own, preventing them from breaking away.
\end{enumerate}
\end{definition}

If the core is non-empty, it indicates that a stable distribution of the total value, $v(\mathcal{N})$, among the players exists such that each player $i$ is satisfied with its allocated utility $x_i$. Conversely, if the core is empty, it implies that the grand coalition cannot be stabilized. In this case, some players have an incentive to break away from the grand coalition and form smaller coalitions.

\begin{definition}
\label{DF:convex:game}
A coalitional game with transferable payoffs $(\mathcal{N}, v)$ is \emph{convex} if \cite{r1}
\begin{equation}
\label{EQ:convexity:definition}
v(S_1) + v(S_2) \leq v(S_1 \cup S_2) + v(S_1 \cap S_2), \quad \forall S_1, S_2 \subseteq \mathcal{N}.
\end{equation}
\end{definition}
We will use the following fact \cite{r1} later in our analysis.
\begin{theorem}
\label{TH:convex:game:has:non:empty:core}
If the game $(\mathcal{N}, v)$ is convex, then its core is non-empty.     
\end{theorem}

\subsection{$\mathbb{D}_c$-Stability}

The concept of $\mathbb{D}_c$-stability \cite{apt2009generic} is used to analyze the stability of coalition structures when players are partitioned into multiple disjoint coalitions. This is relevant when all the players do not cooperate to form a single coalition, but instead form smaller coalitions.

First, we define some terminology \cite{apt2009generic}. Given the set $\mathcal{N}$, a \emph{collection} $\mathbf{S} = \{S_1, \ldots, S_k\}$ is a set of mutually disjoint coalitions of the players in $\mathcal{N}$.  $\mathbf{P} = \{P_1, \ldots, P_n\}$ is a \textit{partition} of the set $\mathcal{N}$ if $P_i \subseteq \mathcal{N}$, $P_i \cap P_j = \emptyset$ for all $P_i, P_j \in \mathbf{P}$ such that $i \neq j$, and $\bigcup_{i=1}^{n} P_i = \mathcal{N}$.  A collection $\mathbf{S} = \{S_1, \dots, S_k\}$ is \textit{$\mathbf{P}-compatible$} if $\bigcup_{i=1}^{k} S_i \subseteq P_j$ for some coalition $P_j \in \mathbf{P}$, and a coalition ${S}$ is \textit{$\mathbf{P}$-incompatible} if $S \not\subseteq P_i$ for every $P_i \in \mathbf{P}$. Given a collection $\mathbf{S} = \{S_1, \ldots, S_k\}$ and a partition $\mathbf{P} = \{P_1, \ldots, P_n\}$, let
\begin{equation}
\mathbf{S}\left[\mathbf{P}\right]=\{\cup_{i=1}^k S_i\cap P_1,\ldots,\cup_{i=1}^k S_i\cap P_n\}.
\end{equation}  
Note that $\mathbf{S}\left[\mathbf{P}\right]$ is the partitioning of the players in $\cup_{i=1}^k S_i$ into coalitions according to the partition $\mathbf{P}$. For a collection $\mathbf{S}=\{S_1,\ldots,S_k\}$, let
\[
v(\mathbf{S})=\sum\limits_{i=1}^kv(S_i)
\]
denote the \emph{value} of collection $\mathbf{S}$.

\begin{definition}
A partition \(\mathbf{P}\) of the set of players \(\mathcal{N}\) is $\mathbb{D}_c$-stable if
  \[
  v(\mathbf{S}[\mathbf{P}]) \geq v(\mathbf{S}), \quad  \forall \mathbf{S}  
 \subseteq \mathcal{N}.
  \]  
\end{definition}

That is, a partition \(\mathbf{P}\) is $\mathbb{D}_c$-stable if for every possible collection  \(\mathbf{S} = \{S_1, \ldots, S_k\}\), the total value when players form coalitions according to \(\mathbf{P}\) is at least as large as when they form coalitions according to \(\mathbf{S}\).

The following result provides a useful necessary and sufficient condition for a partition \(\mathbf{P}\) to be $\mathbb{D}_c$-stable \cite{apt2009generic}.

\begin{theorem}
\label{TH:Dc:stability:conditions}
A partition $\mathbf{P} = \{P_1, \dots, P_n\}$ of $\mathcal{N}$ is $\mathbb{D}_c$-stable if and only if the following two conditions are satisfied:
\begin{enumerate}
    \item For every $\mathbf{P}$-compatible collection $\mathbf{S} = \{S_1, \dots, S_k\}$, the following holds:
    \begin{equation}\label{eq:3}
        v\left(\bigcup_{i=1}^{k} S_i\right) \geq \sum_{i=1}^{k} v(S_i). 
    \end{equation}
    \item For every $\mathbf{P}$-incompatible coalition $S$, the following holds:
    \begin{equation}\label{eq:4}
        \sum_{i=1}^{n} v(S \cap P_i) \geq v(S). 
    \end{equation}
\end{enumerate}
\end{theorem}

Intuitively, the condition in \eqref{eq:3} says that for every subset of players within a coalition \(P_j \in \mathbf{P}\), forming a single coalition is at least as beneficial as splitting into smaller groups. Also, the condition in \eqref{eq:4} says that for the players of any coalition \(S\) that spans multiple coalitions of $\mathbf{P}$, splitting into multiple coalitions according to \(\mathbf{P}\) is at least as beneficial as forming a single coalition \(S\).

\begin{remark}
We now compare the concepts of core and $\mathbb{D}_c$-stability. The concept of core is useful for studying conditions under which it is beneficial for \emph{all} users to cooperate among themselves to form the grand coalition and no subgroup has an incentive to break away. The concept of $\mathbb{D}_c$-stability, on the other hand, is useful for studying conditions under which it is beneficial for users to divide themselves into multiple disjoint coalitions according to some partition $\mathbf{P} = \{P_1, \dots, P_n\}$.
\end{remark}

\section{System Model and Problem Formulation}
\label{SC:system:model:problem:formulation}
\begin{figure}
\centering
\includegraphics[width=0.5\textwidth]{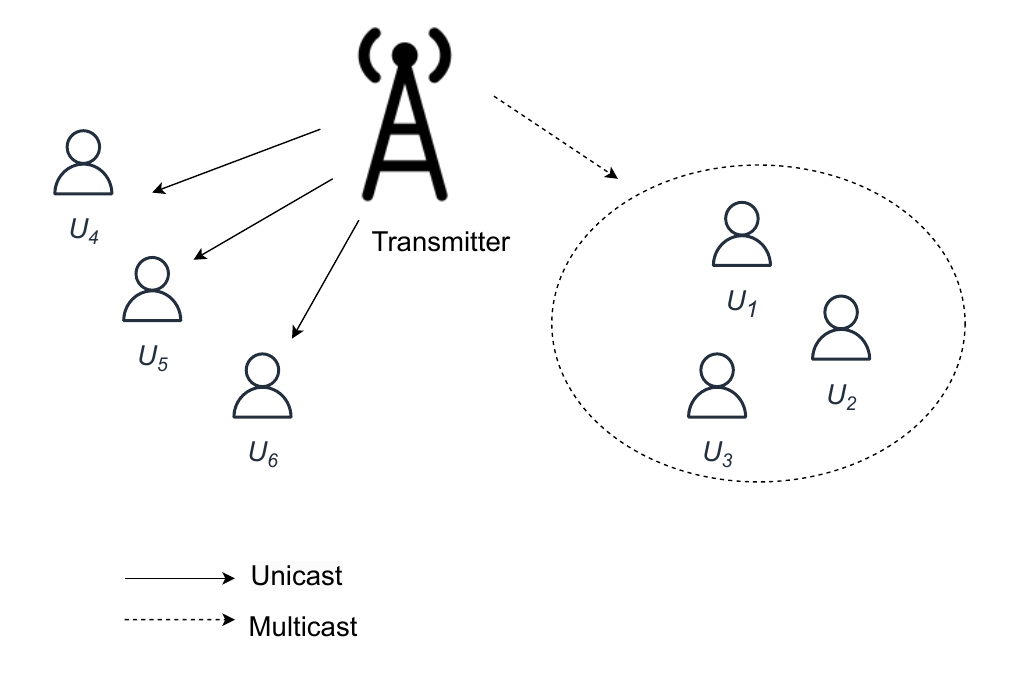}
\caption{The figure shows an example network with $N = 6$ users.}
\label{fig:verticalcell}
\end{figure}

Consider a wireless network in which there is a transmitter and a set $\mathcal{N} = \{1, \ldots, N\}$ of users (players), all of whom want to download a popular file of size $X$ bits from the transmitter (see Fig. \ref{fig:verticalcell}). The channel gain from the transmitter to every user in $\mathcal{N}$ is computed using reference signals and is hence known. Let $R_i$ be the data rate at which data can be transferred from the transmitter to user $i$. The transmitter is assumed to transmit at a fixed power level $P_{Tx}$. User $i \in \mathcal{N}$ obtains a valuation $U_i$ from downloading the file. Let $P_{Rx,i}$ be the power consumed by the device of user $i$ when it downloads data from the transmitter. Suppose the cost incurred at a user (respectively, at the transmitter) when energy $E$ is expended is $aE$ (respectively, $bE$). Also, the cost of using the channel owned by the transmitter  for duration $t$ seconds, i.e., the bandwidth cost, is $wt$. Here, $a$, $b$, and $w$ are parameters whose values are fixed.

Consider a set $S \subseteq \mathcal{N}$ of users who cooperate among themselves and form a coalition. Let
\begin{equation}  
R_S = \min_{i \in S} R_i. 
\end{equation} 
Since the users in $S$ form a coalition, the transmitter multicasts the popular file to those users at the rate $R_S$.  Also, the energy cost and bandwidth cost incurred by the transmitter in distributing the file to the users in $S$  need to be paid by those users. 

The sum of utilities of the users in $S$ is given by
\begin{equation} 
\label{EQ:value:function}
v(S) = \sum_{i \in S} U_i - a \sum_{i \in S} \frac{P_{Rx,i}X}{R_S} - \frac{b P_{Tx} X}{R_S} - \frac{wX}{R_S},
\end{equation}
where $v(S)$ is the value of the coalition $S$ and $\sum_{i \in S} U_i$  represents the total valuation or benefit that all users in the coalition $S$ derive from the popular file. Note that the amount of time required to download the file is $\frac{X}{R_S}$; therefore, the energy consumed by user $i$ in receiving the file is $\frac{P_{Rx,i}X}{R_S}$. Hence,   $a \sum_{i \in S} \frac{P_{Rx,i}X}{R_S}$ is the total energy cost incurred by all users in $S$ to receive the file. Similarly, \textbf{$\frac{b P_{Tx} X}{R_S}$} is the energy cost incurred by the transmitter in transmitting the file to the users in $S$, and \textbf{$\frac{wX}{R_S}$} is the bandwidth cost of transmitting the file.   

Our goal is to find conditions under which it is beneficial for the users to cooperate with each other. Specifically, we study two cases: 1) when the set of \emph{all} users in $\mathcal{N}$ cooperate among themselves, and 2) when the set of users is partitioned into multiple disjoint coalitions such that the users of every
coalition cooperate among themselves. In particular, our objectives are as follows:
\begin{enumerate}[(a)]
    \item 
    To identify conditions under which the core is non-empty and those under which it is empty.
    \item 
    To identify conditions under which a given partition \(\mathbf{P}\) of $\mathcal{N}$ is $\mathbb{D}_c$-stable.  
\end{enumerate}

We now introduce some notation for later use. Let $\alpha_i = a P_{Rx,i} X$, $\beta = b P_{Tx} X$, and $\gamma = wX$. Then, we can rewrite \eqref{EQ:value:function} as
\begin{equation} 
\label{EQ:value:function:simplified}
v(S) = \sum_{i \in S} U_i - \sum_{i \in S} \frac{\alpha_i}{R_S} - \frac{(\beta+\gamma)}{R_S}.
\end{equation}

\section{Networks with Non-Empty Core}
\label{SC:core:non:empty}
First, we consider the symmetric case of the game described in Section \ref{SC:system:model:problem:formulation} where every user in the
network has the same data rate, say $R_0$, and the same receive power, say $P_{Rx}$. 
\begin{theorem}
\label{TH:symmetric:case:non:empty:core}
Let  $R_i = R_0$  and $P_{Rx,i} = P_{Rx}$, $\forall i \in \mathcal{N}$. Then the game has a non-empty core.  
\end{theorem}

Intuitively, since all users have the same date rate and receive power, the file can be efficiently distributed by multicasting it to all users in $\mathcal{N}$ at the rate $R_0$, which is the same rate at which the file would need to be transmitted if it were to be unicast to a single user or multicast to a subset of users in $\mathcal{N}$. Also, when all users in $\mathcal{N}$ cooperate, the energy cost and bandwidth cost incurred by the transmitter can be shared among all of them and the burden on an individual user can be low. Hence, users have an incentive to cooperate among themselves and form the grand coalition, due to which the core is non-empty. 

\begin{IEEEproof}[Proof of Theorem \ref{TH:symmetric:case:non:empty:core}]
Note that \( R_{\mathcal{N}} = \min_{i \in \mathcal{N}} R_i = R_0 \). Also,  \( \alpha_1 = \ldots = \alpha_N  = a P_{Rx} X =\alpha \mbox{ (say).}\)
By \eqref{EQ:value:function:simplified},
\begin{equation}
\label{EQ:value:function:nec:sc}    
v(S) = \sum_{i \in S} U_i - \frac{\alpha |S|}{R_0} - \frac{\left(\beta  + \gamma\right)}{R_0}, \, \forall S \subseteq \mathcal{N}.
\end{equation}
By \eqref{EQ:value:function:nec:sc}, we get that for any two coalitions \( S_1 \) and \( S_2 \),
\begin{align}
&v(S_1) + v(S_2) \nonumber \\
&= \sum_{i \in S_1} U_i + \sum_{i \in S_2} U_i
 - \frac{\alpha (|S_1| + |S_2|)}{R_0}  - \frac{2\left(\beta  + \gamma\right)}{R_0}, \label{EQ:vS1:p:vS2} \\
& \mbox{and } v(S_1 \cup S_2) + v(S_1 \cap S_2) \nonumber \\
&= \sum_{i \in S_1 \cup S_2} U_i + \sum_{i \in S_1 \cap S_2} U_i \nonumber \\
& \quad- \frac{\alpha (|S_1 \cup S_2| + |S_1 \cap S_2|)}{R_0}  - \frac{2\left(\beta  + \gamma\right)}{R_0}. \label{EQ:vS1uS2:p:vS1iS2}
\end{align}
Also, note that
\begin{align}
\sum_{i \in S_1 \cup S_2} U_i &= \sum_{i \in S_1} U_i + \sum_{i \in S_2} U_i - \sum_{i \in S_1 \cap S_2} U_i, \label{EQ:sums:of:utilities} \\
|S_1 \cup S_2| &= |S_1| + |S_2| - |S_1 \cap S_2|. \label{EQ:S1:u:S2}
\end{align}
By \eqref{EQ:vS1:p:vS2}, \eqref{EQ:vS1uS2:p:vS1iS2}, \eqref{EQ:sums:of:utilities}, and \eqref{EQ:S1:u:S2}, we get
\begin{align*}
&v(S_1 \cup S_2) + v(S_1 \cap S_2) \\
&= \left( \sum_{i \in S_1} U_i + \sum_{i \in S_2} U_i - \sum_{i \in S_1 \cap S_2} U_i \right) \\
&\quad + \sum_{i \in S_1 \cap S_2} U_i - \frac{\alpha \left( |S_1| + |S_2| \right)}{R_0}  - \frac{2\left(\beta  + \gamma\right)}{R_0} \\
&= \sum_{i \in S_1} U_i + \sum_{i \in S_2} U_i 
- \frac{\alpha \left( |S_1| + |S_2| \right)}{R_0} - \frac{2\left(\beta  + \gamma\right)}{R_0} \\
&= v(S_1) + v(S_2).
\end{align*}
So, by Definition \ref{DF:convex:game}, the game is convex. Hence, by Theorem \ref{TH:convex:game:has:non:empty:core}, it has a  non-empty core. The result follows.
\end{IEEEproof}

Next, the following proposition explicitly identifies a payoff profile of the $N$ players, which lies in the core. 

\begin{proposition}
\label{PN:symmetric:case:payoff:profile}
Let
\begin{equation}
\label{EQ:symmetric:case:payoff:profile}
\ x_i = U_i -  \frac{\alpha}{R_0} - \frac{(\beta + \gamma)}{N R_0}, \quad \forall i \in \mathcal{N}.
\end{equation} 
Then the payoff profile \( (x_1, x_2, \ldots, x_N) \) lies in the core. 
\end{proposition}

Note that the payoff profile in \eqref{EQ:symmetric:case:payoff:profile} is achieved when each user in $\mathcal{N}$ makes the same payment, $\frac{(\beta + \gamma)}{NR_0}$, towards the total transmission and bandwidth cost incurred, $\frac{(\beta + \gamma)}{R_0}$.

\begin{IEEEproof}[Proof of Proposition \ref{PN:symmetric:case:payoff:profile}]
By \eqref{EQ:symmetric:case:payoff:profile} and \eqref{EQ:value:function:nec:sc},
\begin{equation}
\label{EQ:sc:pp:efficiency}
    \sum_{i \in \mathcal{N}} x_i = \sum_{i \in \mathcal{N}} U_i - \frac{N\alpha}{R_0} - \frac{(\beta + \gamma)}{R_0} = v(\mathcal{N}).
\end{equation}
Also, by \eqref{EQ:symmetric:case:payoff:profile}, for every $S \subseteq \mathcal{N}$,
\begin{eqnarray}
    \sum_{i \in S} x_i & = & \sum_{i \in S} U_i - \frac{|S|\alpha}{R_0} - \frac{|S|(\beta + \gamma)}{NR_0} \nonumber \\
                       & \geq & \sum_{i \in S} U_i - \frac{|S|\alpha}{R_0} - \frac{N(\beta + \gamma)}{NR_0} \quad \mbox{(since } |S| \leq N) \nonumber \\
                       & = & \sum_{i \in S} U_i - \frac{|S|\alpha}{R_0} - \frac{(\beta + \gamma)}{R_0} \nonumber \\
                       & = & v(S) \quad \mbox{(by \eqref{EQ:value:function:nec:sc})}. \label{EQ:sc:pp:coalitional:rationality}
\end{eqnarray}

By \eqref{EQ:sc:pp:efficiency} and \eqref{EQ:sc:pp:coalitional:rationality}, the payoff profile \( (x_1, x_2, \ldots, x_N) \) satisfies the properties of efficiency and coalitional rationality in Definition \ref{DF:core}. So, by definition, it lies in the core. The result follows. 
\end{IEEEproof}

Next, consider the game described in Section \ref{SC:system:model:problem:formulation} and let $R_{min} =\min_{i \in \mathcal{N}} R_i$, $R_{max} =\max_{i \in \mathcal{N}} R_i$, $\alpha_{min} =\min_{i \in \mathcal{N}} \alpha_i$, and $\alpha_{max} =\max_{i \in \mathcal{N}} \alpha_i$. The following result establishes a sufficient condition under which the core is non-empty.  

\begin{theorem}
\label{TH:Rmax:by:Rmin:ub}
The core is non-empty if
\begin{equation}
\label{EQ:Rmax:by:Rmin:ub}
\frac{R_{max}}{R_{min}} \leq \left( \frac{N}{N-1} \right) \left( \frac{\alpha_{min}(N-1) + \beta + \gamma}{\alpha_{max}N+\beta+\gamma} \right).
\end{equation}
\end{theorem}

Intuitively, Theorem \ref{TH:Rmax:by:Rmin:ub} says that when all users $i \in \mathcal{N}$ have similar data rates $R_i$ (which is the case when $\frac{R_{max}}{R_{min}}$ is sufficiently small), it is beneficial for them to cooperate to form the grand coalition. This is because users with high rates experience only a small reduction in the data rate at which they receive the file from the transmitter when they cooperate with users with low rates to form the grand coalition, relative to the case where they break away from the grand coalition to form a separate coalition. Also, it is beneficial for all users to cooperate and form the grand coalition because in this case, the transmission and bandwidth costs are shared among all users, and hence an individual user has to make only a small payment towards them to the transmitter.   However, as shown in Section \ref{SC:core:empty}, if rate disparities grow too large, high-rate users may prefer to break away from the grand coalition and form smaller groups which can download the file at high rates from the transmitter. Theorem \ref{TH:Rmax:by:Rmin:ub} provides a threshold for the level of disparity, ensuring that as long as the ratio of maximum to minimum rates, $\frac{R_{max}}{R_{min}}$, stays below this threshold, the grand coalition remains stable.

\begin{IEEEproof}[Proof of Theorem \ref{TH:Rmax:by:Rmin:ub}]
From \eqref{EQ:value:function:simplified}, we get
\begin{eqnarray*}
v(\mathcal{N})  & = & \sum_{i \in \mathcal{N}} U_i -  \sum_{i \in \mathcal{N}} \frac{\alpha_{i}}{R_{min}} - \frac{\beta+\gamma}{R_{min}}. \\
& \geq & \sum_{i \in \mathcal{N}} U_i - \frac{N\alpha_{max}}{R_{min}} - \frac{\beta+\gamma}{R_{min}}. 
\end{eqnarray*}
Hence, there exist $x_i, \, i \in \mathcal{N}$, such that $\sum_{i \in \mathcal{N}} x_i = v(\mathcal{N})$, and
\begin{equation}
\label{EQ:Rmax:by:Rmin:ub:xi:ub}
x_i \geq U_i - \frac{(\alpha_{max} N + \beta+\gamma)}{N R_{min}}, \quad \forall i \in \mathcal{N}.
\end{equation}

Now, let $S \subseteq \mathcal{N}$ be a subset of $\mathcal{N}$ such that $S \neq \mathcal{N}$. Then, by  \eqref{EQ:Rmax:by:Rmin:ub:xi:ub},
\begin{equation}
\label{EQ:sum:xi:lb}
\sum_{i \in S} x_i \geq \sum_{i \in S} U_i - \frac{|S|(\alpha_{max} N + \beta+\gamma)}{N R_{min}}. 
\end{equation}
Also, by \eqref{EQ:value:function:simplified} and the fact that $\alpha_i \geq \alpha_{min}$ and $R_i \leq R_{max}$ $\forall i \in \mathcal{N}$, we get
\begin{equation}
\label{EQ:vS:ub}
v(S)   \leq \sum_{i \in S} U_i - \frac{(\alpha_{min}|S| +\beta+\gamma)}{R_{max}}. 
\end{equation}

By (\ref{EQ:sum:xi:lb}) and (\ref{EQ:vS:ub}), $\sum_{i \in S} x_i \geq v(S)$ if
\[
\sum_{i \in S} U_i - \frac{|S|(\alpha_{max} N + \beta+\gamma)}{N R_{min}} \geq \sum_{i \in S} U_i - \frac{(\alpha_{min}|S| +\beta+\gamma)}{R_{max}}.
\]
The preceding inequality is equivalent to
\[
\frac{R_{max}}{R_{min}} \leq \left( \frac{N}{|S|} \right) \left( \frac{\alpha_{min}|S| + \beta+\gamma}{\alpha_{max}N+\beta+\gamma} \right).
\]
This inequality holds if (\ref{EQ:Rmax:by:Rmin:ub}) holds since $|S| \leq N-1$. The result follows. 
\end{IEEEproof}

\section{Networks with Empty Core}
\label{SC:core:empty}
\subsection{Given Minimum and Second Minimum Data Rates}
Let \( k = \underset{i \in \mathcal{N}}{\operatorname{argmin}}  (R_i) \), \( \mathcal{N}' = \mathcal{N} \setminus \{k\} \), and \( j = \underset{i \in \mathcal{N}'}{\operatorname{argmin}}  (R_i) = \lambda R_k \), where \(\lambda \geq 1\). Hence, the minimum and second minimum data rates are $R_k$ and $R_j = \lambda R_k$, respectively.
\begin{theorem}
\label{TH:min:second:min:core:empty}
Suppose the reception power is the same for all users: $P_{Rx,i} = P_{Rx}$, $\forall i \in \mathcal{N}$. Also, let $R_k$ and $R_j = \lambda R_k$ be the minimum and second minimum data rates, respectively, where $\lambda \geq 1$. The core is empty if
\begin{equation}
\label{EQ:min:second:min:data:rates:core:empty}
\lambda  > 1 +\frac{\beta+\gamma}{\alpha (N - 1)}. 
\end{equation}
\end{theorem}

Intuitively, since the data rate at which the users of the grand coalition download the file equals the minimum rate $R_k$, when $\lambda$ is high, the large gap between $R_k$ and the second minimum rate, $R_j$,  gives the  $N-1$ users in \( \mathcal{N}' \) incentive to separate from $k$ and operate at rate $R_j$. Condition \eqref{EQ:min:second:min:data:rates:core:empty} specifies the threshold at which the cost of remaining with the user, $k$, with the minimum data rate outweighs the benefits of cooperation, leading to an empty core.

\begin{IEEEproof}[Proof of Theorem \ref{TH:min:second:min:core:empty}]
Suppose \eqref{EQ:min:second:min:data:rates:core:empty} holds. To reach a contradiction, assume that the core is non-empty. Then there exist \( \{x_1, x_2, \ldots, x_{N}\} \) such that \eqref{EQ:core:efficiency} and \eqref{EQ:core:coalitional:rationality} hold.
Hence, the following two inequalities hold:
\begin{equation}
\label{EQ:min:second:min:data:rates:core:empty:two:inequalities}
\sum_{i \in \mathcal{N}'} x_i \geq v(\mathcal{N}') \quad \text{and} \quad x_k \geq v(\{k\}).
\end{equation}
Adding the two inequalities in \eqref{EQ:min:second:min:data:rates:core:empty:two:inequalities}, we get
\begin{equation}
\label{EQ:min:second:min:data:rates:core:empty:sum:of:two:inequalities}    
\sum_{i \in \mathcal{N}} x_i \geq v(\mathcal{N}') + v(\{k\}).
\end{equation}
By  \eqref{EQ:min:second:min:data:rates:core:empty:sum:of:two:inequalities} and \eqref{EQ:core:efficiency}, we get
\begin{equation}
\label{EQ:min:second:min:data:rates:core:empty:vN:geq:vNprime:vk}    
    v(\mathcal{N}) \geq v(\mathcal{N}') + v(\{k\}).
\end{equation}

Now, consider
\begin{align}
&v(\mathcal{N}') + v(\{k\}) - v(\mathcal{N}) \notag \\
&=  \left( \sum_{i \in \mathcal{N}'} U_i - \frac{\alpha |\mathcal{N}'|}{R_j} - \frac{\beta+\gamma}{R_j} \right) +
 \left( U_k - \frac{\alpha}{R_k} - \frac{\beta+\gamma}{R_k} \right) \notag \\
 & \quad- \left(\sum_{i \in \mathcal{N}} U_i - \frac{\alpha |\mathcal{N}|}{R_k} - \frac{\beta+\gamma}{R_k} \right) \notag \\
 &= \frac{\alpha (N - 1)}{R_k}  \left(1 - \frac{1}{\lambda}\right)  - \frac{\beta + \gamma}{\lambda R_k} \label{EQ:min:second:min:data:rates:core:empty:difference:eq1}    \\
 &=\frac{1}{\lambda R_k} \left\{ \alpha (N - 1)(\lambda-1)-(\beta+\gamma) \right\} \notag \\
 &> 0, \label{EQ:min:second:min:data:rates:core:empty:difference:eq2}
\end{align}
where \eqref{EQ:min:second:min:data:rates:core:empty:difference:eq1} follows from the facts that $\mathcal{N} = \mathcal{N}'\cup\{k\}$, $|\mathcal{N}| = N$, $|\mathcal{N}'|=N-1$, and $R_j = \lambda R_k$. Also, the inequality in \eqref{EQ:min:second:min:data:rates:core:empty:difference:eq2} follows from \eqref{EQ:min:second:min:data:rates:core:empty}.  Thus, we have proved that $v(\mathcal{N}') + v(\{k\}) - v(\mathcal{N}) > 0$, which contradicts  \eqref{EQ:min:second:min:data:rates:core:empty:vN:geq:vNprime:vk}. This shows that the core is empty.  
\end{IEEEproof}

\subsection{Given Minimum and Maximum Data Rates}
Let \( m = \underset{i \in \mathcal{N}}{\operatorname{argmax}}  (R_i) \), so that the maximum data rate is $R_m$. Also, let $R_m = \mu R_k$, where $\mu \geq 1$, and $\tilde{\mathcal{N}}=\mathcal{N} \backslash\{m\}$. 
\begin{theorem}
\label{TH:min:max:core:empty}
Suppose the reception power is the same for all users: $P_{Rx,i} = P_{Rx}$, $\forall i \in \mathcal{N}$. Also, let $R_k$ and $R_m = \mu R_k$ be the minimum and maximum data rates, respectively, where $\mu \geq 1$. The core is empty if
\begin{equation}
\label{EQ:min:max:core:empty:mu:inequality}
\mu  > 1 + \frac{\beta+\gamma}{\alpha}.
\end{equation}
\end{theorem}

Intuitively, when $\mu$ is large, the user, $m$, with the maximum rate has a much higher rate than $R_k$, which is the rate at which the users of the grand coalition download the file. Hence, operating at $R_k$ leads to significant inefficiency for user $m$. Inequality \eqref{EQ:min:max:core:empty:mu:inequality} gives the threshold at which the  user, $m$, with the maximum rate prefers to break away, and the grand coalition becomes unstable, making the core empty.

\begin{IEEEproof}[Proof of Theorem \ref{TH:min:max:core:empty}] 
Suppose \eqref{EQ:min:max:core:empty:mu:inequality} holds. To reach a contradiction, assume that the core is non-empty. Similar to the derivation of \eqref{EQ:min:second:min:data:rates:core:empty:vN:geq:vNprime:vk}, we get
\begin{equation}
\label{EQ:min:max:core:empty:vN:geq:vNprime:vk}    
    v(\mathcal{N}) \geq v(\tilde{\mathcal{N}}) + v(\{m\}).
\end{equation}

Now, consider
\begin{align}
&v(\tilde{\mathcal{N}}) + v(\{m\}) - v(\mathcal{N}) \notag \\
&=  \left( \sum_{i \in \tilde{\mathcal{N}}} U_i - \frac{\alpha |\tilde{\mathcal{N}}|}{R_k} - \frac{\beta+\gamma}{R_k} \right) +
 \left( U_m - \frac{\alpha}{R_m} - \frac{\beta+\gamma}{R_m} \right) \notag \\
 & \quad- \left(\sum_{i \in \mathcal{N}} U_i - \frac{\alpha |\mathcal{N}|}{R_k} - \frac{\beta+\gamma}{R_k} \right) \notag \\
 &=\frac{\alpha}{ R_k} \left\{ 1- \frac{1}{\mu \alpha} (\alpha+\beta+\gamma) \right\} \label{EQ:min:max:core:empty:difference:eq1} \\
 &> 0, \label{EQ:min:max:core:empty:difference:eq2}
\end{align}
where \eqref{EQ:min:max:core:empty:difference:eq1} follows from the facts that $\mathcal{N} = \tilde{\mathcal{N}}\cup\{m\}$, $|\mathcal{N}| = N$, $|\tilde{\mathcal{N}}|=N-1$, and $R_m = \mu R_k$. Also, the inequality in \eqref{EQ:min:max:core:empty:difference:eq2} follows from \eqref{EQ:min:max:core:empty:mu:inequality}.  Thus, we have proved that $v(\tilde{\mathcal{N}}) + v(\{m\}) - v(\mathcal{N}) > 0$, which contradicts  \eqref{EQ:min:max:core:empty:vN:geq:vNprime:vk}. This shows that the core is empty.  
\end{IEEEproof}

\section{Conditions for a Partition to be $\mathbb{D}_c$-Stable}
\label{SC:Dc:stability}
\subsection{General Result}
Consider the game described in Section \ref{SC:system:model:problem:formulation} and let $\mathbf{P} = \{P_1, \ldots, P_n\}$ be a partition of the set $\mathcal{N}$. In this section, we study conditions under which the partition $\mathbf{P}$ is  $\mathbb{D}_c$-stable. 

For $i \in \{1, \ldots, n\}$, let $|P_i| = N_i \geq 1$, i.e., there are $N_i$ users in the coalition $P_i$. 
For $i \in \{1, \ldots, n\}$ and $j \in \{1, \ldots, N_i\}$, let \( R_{i,j} \) denote the rate of the \( j^{\text{th}} \) user belonging to $P_i$.  Let  \( R_{i,\text{max}} \) and \( R_{i,\text{min}} \) denote the maximum and minimum rate, respectively, of the users of $P_i$. Hence, $R_{i,j} \in [R_{i,min}, R_{i,max}]$, $\forall j \in \{1, \ldots, N_i\}$. We assume that
    \small
    \begin{equation}
        R_{1,\text{min}} \leq R_{1,\text{max}} \leq R_{2,\text{min}} \leq R_{2,\text{max}} \leq \ldots \leq R_{n,\text{min}} 
        \leq R_{n,\text{max}}.
    \end{equation}
    \normalsize
Let  \( P_{Rx,\text{max}} \) and \( P_{Rx,\text{min}} \) denote the maximum and minimum receive power, respectively, of each user of $\mathcal{N}$. Also, let $\alpha_{max}=aX P_{Rx,\text{max}}$ and $\alpha_{min}=aX P_{Rx,\text{min}}$. 

\begin{theorem}
\label{TH:Dc:stability:sufficient:conditions}
The partition $\mathbf{P} = \{P_1, \dots, P_n\}$ of $\mathcal{N}$ is $\mathbb{D}_c$-stable if    
\begin{equation}
\label{EQ:P:incompatible:sufficient:condition}
    \frac{R_{i+1,min}}{R_{i,min}} \geq \frac{\alpha_{min} + \beta + \gamma}{\alpha_{min}}, \quad \forall i \in \{1, \ldots,n-1\},
\end{equation}
and
\begin{equation}
\label{EQ:P:compatible:sufficient:condition}
    \frac{R_{j,max}}{R_{j,min}} \leq \frac{2 \left(\alpha_{min}  + \beta + \gamma \right)}{\alpha_{max} \left| P_j \right| + \beta + \gamma}, \quad \forall j \in \{1, \ldots, n\}.
\end{equation}
\end{theorem}

Intuitively, condition \eqref{EQ:P:incompatible:sufficient:condition} ensures that users in different coalitions, $P_i$ and $P_{i^{\prime}}$, do not benefit from merging, as the data rates of users of different coalitions are significantly different. On the other hand, condition \eqref{EQ:P:compatible:sufficient:condition} prevents users in the same coalition $P_j$ from splitting into smaller groups, since the data rates of all users within $P_j$ are similar. When combined, these two conditions guarantee that every user remains in its own coalition, which results in $\mathbb{D}_c$-stability.

We now prove two lemmas (Lemmas \ref{LM:P:incompatible:sufficient:condition} and \ref{LM:P:compatible:sufficient:condition}), using which we will later prove Theorem \ref{TH:Dc:stability:sufficient:conditions}.

\begin{lemma}
\label{LM:P:incompatible:sufficient:condition}
The condition in \eqref{eq:4} is satisfied if \eqref{EQ:P:incompatible:sufficient:condition} holds. 
\end{lemma}
\begin{IEEEproof}
Let ${S}$ be a $\mathbf{P}$-incompatible coalition. Then by definition, there exist an integer $n_S \geq 2$ and a set of indices $\{i_1, \ldots, i_{n_S}\}$ such that $S \cap P_i \neq \emptyset$ for $i \in \{i_1, \ldots, i_{n_S}\}$ and $S \cap P_i = \emptyset$ otherwise. 

Consider
\begin{align}
 &\sum_{i=1}^n v(S \cap P_i) - v(S)   \notag \\
 &= \sum_{j=1}^{n_S} v(S \cap P_{i_j}) - v(S) \notag \\
 &= \sum_{j=1}^{n_S} \left\{ \sum_{l \in S \cap P_{i_j}} U_l - \sum_{l \in S \cap P_{i_j}} \frac{\alpha_l}{R_{S \cap P_{i_j}}} - \frac{\beta + \gamma}{R_{S \cap P_{i_j}}} \right\} \notag \\
& \quad - \left\{ \sum_{i \in S} U_i - \sum_{i \in S} \frac{\alpha_i}{R_S} - \frac{\beta + \gamma}{R_S} \right\} \label{EQ:P:incompatible:sufficient:condition:proof:eq1} \\
&= \sum_{j=2}^{n_S} \left\{ \sum_{l \in S \cap P_{i_j}} \alpha_l \left( \frac{1}{R_{S \cap P_{i_1}}} - \frac{1}{R_{S \cap P_{i_j}}} \right) - \frac{\beta + \gamma}{R_{S \cap P_{i_j}}}
\right\} \label{EQ:P:incompatible:sufficient:condition:proof:eq2} \\
&\geq \sum_{j=2}^{n_S} \left\{ \sum_{l \in S \cap P_{i_j}} \alpha_{min} \left( \frac{1}{R_{S \cap P_{i_1}}} - \frac{1}{R_{S \cap P_{i_j}}} \right) - \frac{\beta + \gamma}{R_{S \cap P_{i_j}}}
\right\} \label{EQ:P:incompatible:sufficient:condition:proof:eq3}  \\
&\geq \sum_{j=2}^{n_S} \left\{  \alpha_{min} \left( \frac{1}{R_{S \cap P_{i_1}}} - \frac{1}{R_{S \cap P_{i_j}}} \right) - \frac{\beta + \gamma}{R_{S \cap P_{i_j}}}
\right\} \label{EQ:P:incompatible:sufficient:condition:proof:eq4} \\
&\geq \sum_{j=2}^{n_S} \left\{  \alpha_{min} \left( \frac{1}{R_{i_1,max}} \right) - \frac{1}{R_{i_j,min}} \left( \alpha_{min} +\beta + \gamma \right) \right\} \label{EQ:P:incompatible:sufficient:condition:proof:eq5} \\
&= \sum_{j=2}^{n_S}  \frac{\alpha_{min}}{R_{i_j,min}} \left\{ \frac{R_{i_j,min}}{R_{i_1,max}} - \left( \frac{\alpha_{min} +\beta + \gamma}{\alpha_{min}} \right)
\right\} \notag \\
& \geq 0, \label{EQ:P:incompatible:sufficient:condition:proof:eq6}
\end{align}
where \eqref{EQ:P:incompatible:sufficient:condition:proof:eq1} follows from \eqref{EQ:value:function:simplified}. Also, \eqref{EQ:P:incompatible:sufficient:condition:proof:eq2} follows from the following facts: (i) $\sum_{j=1}^{n_S}  \sum_{l \in S \cap P_{i_j}} U_l = \sum_{i \in S} U_i$, due to which the utility terms in \eqref{EQ:P:incompatible:sufficient:condition:proof:eq1} cancel, (ii)  $R_S = R_{S \cap P_{i_1}}$, and (iii) $\sum_{i \in S} \frac{\alpha_i}{R_S} = \sum_{j=1}^{n_S}  \sum_{l \in S \cap P_{i_j}} \frac{\alpha_l}{R_{S \cap P_{i_1}}}$.
Equation \eqref{EQ:P:incompatible:sufficient:condition:proof:eq3} follows from the facts that $\alpha_l \geq \alpha_{min}$, $\forall l \in S$, and $\frac{1}{R_{S \cap P_{i_1}}} - \frac{1}{R_{S \cap P_{i_j}}} \geq 0$, $\forall j \in \{2, \ldots, n_S\}$. Equation \eqref{EQ:P:incompatible:sufficient:condition:proof:eq4} follows from the fact that $|S \cap P_{i_j}| \geq 1$, $\forall j \in \{2, \ldots, n_S\}$. Equation \eqref{EQ:P:incompatible:sufficient:condition:proof:eq5} follows from the fact that $R_{i,j} \in [R_{i,min}, R_{i,max}]$, $\forall j \in \{1, \ldots, N_i\}$. Finally, \eqref{EQ:P:incompatible:sufficient:condition:proof:eq6} follows from \eqref{EQ:P:incompatible:sufficient:condition}.

Thus, $\sum_{i=1}^n v(S \cap P_i) \geq v(S)$, and hence \eqref{eq:4} is satisfied.
\end{IEEEproof}

Let $\mathbf{S} = \{S_1, \dots, S_k\}$ be a $\mathbf{P}$-compatible collection such that $\bigcup_{i=1}^k S_i \subseteq P_j$. Suppose $|S_i| = m_i$, $\forall i \in \{1, \ldots, k\}$. Let $S = \bigcup_{i=1}^k S_i$. 

\begin{lemma}
\label{LM:P:compatible:sufficient:condition}
The condition in \eqref{eq:3} is satisfied if \eqref{EQ:P:compatible:sufficient:condition} holds.
\end{lemma}

\begin{IEEEproof}
Suppose \eqref{EQ:P:compatible:sufficient:condition} holds.
The LHS and RHS of \eqref{eq:3} satisfy the following:
\begin{align}
LHS &= \sum_{i \in S} U_i - \sum_{i \in S} \frac{\alpha_i}{R_{S}} - \frac{(\beta+\gamma)}{R_{S}} \notag \\
&\geq \sum_{i \in S} U_i - \sum_{i \in S} \frac{\alpha_{max}}{R_{j,min}} - \frac{(\beta+\gamma)}{R_{j,min}} \notag \\
&= \sum_{i \in S} U_i - \frac{\alpha_{max}\left( \sum_{i=1}^k m_i\right)}{R_{j,min}} - \frac{(\beta+\gamma)}{R_{j,min}},
\label{EQ:P:compatible:sufficient:condition:eq1}
\end{align}
and
\begin{align}
RHS &= \sum_{i=1}^k \left\{ \sum_{l \in S_i} U_l - \sum_{l \in S_i} \frac{\alpha_l}{R_{S_i}} - \frac{(\beta + \gamma)}{R_{S_i}} 
\right\} \notag \\
& \leq \sum_{i \in S} U_i -  \sum_{i \in S} \frac{\alpha_{min}}{R_{j,max}} - \frac{k(\beta + \gamma)}{R_{j,max}} \notag \\
&= \sum_{i \in S} U_i - \frac{\alpha_{min}\left( \sum_{i=1}^k m_i\right)}{R_{j,max}} - \frac{k(\beta + \gamma)}{R_{j,max}}.
\label{EQ:P:compatible:sufficient:condition:eq2}
\end{align}

By \eqref{EQ:P:compatible:sufficient:condition:eq1} and \eqref{EQ:P:compatible:sufficient:condition:eq2}, we get that $LHS  \geq RHS$ if
\begin{align}
&\sum_{i \in S} U_i - \frac{\alpha_{max}\left( \sum_{i=1}^k m_i\right)}{R_{j,min}} - \frac{(\beta+\gamma)}{R_{j,min}}    \notag \\
& \geq \sum_{i \in S} U_i - \frac{\alpha_{min}\left( \sum_{i=1}^k m_i\right)}{R_{j,max}} - \frac{k(\beta + \gamma)}{R_{j,max}}. \label{EQ:P:compatible:sufficient:condition:eq3}
\end{align}
By algebraic simplification, we get that \eqref{EQ:P:compatible:sufficient:condition:eq3} holds iff the following inequality holds:
\begin{equation}
    \frac{R_{j,max}}{R_{j,min}} \leq \frac{\alpha_{min} \left( \sum_{i=1}^{k} m_i \right) + k(\beta + \gamma)}{\alpha_{max} \left( \sum_{i=1}^{k} m_i \right) + \beta + \gamma}, \quad \forall j \in \{1, \ldots, n\}.  
    \label{EQ:P:compatible:sufficient:condition:eq4}
\end{equation}
However, \eqref{EQ:P:compatible:sufficient:condition:eq4} follows from \eqref{EQ:P:compatible:sufficient:condition} and the fact that $k \geq 2$, $m_i \geq 1$ for all $i \in \{1, \ldots, k\}$, and $\sum_{i=1}^{k} m_i \leq \left| P_j \right|$. Thus, $LHS  \geq RHS$ and hence \eqref{eq:3} is satisfied. 
\end{IEEEproof}

\begin{IEEEproof}[Proof of Theorem \ref{TH:Dc:stability:sufficient:conditions}]
 The result follows from Theorem \ref{TH:Dc:stability:conditions} and Lemmas \ref{LM:P:incompatible:sufficient:condition} and \ref{LM:P:compatible:sufficient:condition}.  
\end{IEEEproof}

\subsection{A Special Case}
Now we shall find a set of sufficient conditions under which no two users cooperate, i.e., each user separately downloads the file from the transmitter. In other words, if  $P_i = \{i\}$, then we will find a set of sufficient conditions under which the partition $\mathbf{P} = \{P_1, \ldots, P_n\}$  is $\mathbb{D}_c$-stable.

\begin{theorem}
\label{TH:singleton:Dc:stability:sufficient:condition}
The partition $\mathbf{P} = \{P_1, \ldots, P_n\}$ is $\mathbb{D}_c$-stable if
\begin{equation}
\label{EQ:singleton:Dc:stability:sufficient:condition}
    \frac{R_{i+1}}{R_i} \geq \frac{\alpha_{i+1} + \beta + \gamma}{\alpha_{i+1}}, \, \forall i \in \{1, \ldots, n-1\}.
\end{equation}
\end{theorem}

Intuitively, the condition in \eqref{EQ:singleton:Dc:stability:sufficient:condition} guarantees that any benefits of cooperation between two different users $i$ and $i^{\prime}$ will be outweighed by the disadvantages due to the large rate difference between the users.  Specifically, whenever two users cooperate, the user with the smaller data rate lowers the  rate at which data will be sent from the transmitter, which increases the  energy and bandwidth costs for the user with the higher rate.  This discourages the formation of any coalition, ensuring $\mathbb{D}_c$-stability of the partition $\mathbf{P}$ comprised of singleton coalitions. Hence, in this case, separately downloading the file via unicast is the most advantageous approach for users.

\begin{IEEEproof}[Proof of Theorem \ref{TH:singleton:Dc:stability:sufficient:condition}]
Suppose \eqref{EQ:singleton:Dc:stability:sufficient:condition} holds. Note that the only $\mathbf{P}$-compatible collections $\mathbf{S} = \{S_1, \dots, S_k\}$ are those for which $\bigcup_{i=1}^{k} S_i$ is a singleton. For such a collection, the inequality \eqref{eq:3} is trivially satisfied. So, to prove the $\mathbb{D}_c$-stability of $\mathbf{P}$, it suffices to show that \eqref{eq:4} is satisfied for every $\mathbf{P}$-incompatible coalition $S$. For such a coalition $S$, similar to the derivation of \eqref{EQ:P:incompatible:sufficient:condition:proof:eq2}, we get
\begin{align}
&\sum_{i=1}^n v(S \cap P_i) - v(S)   \notag \\
&= \sum_{j=2}^{n_S} \left\{ \alpha_{i_j} \left( \frac{1}{R_{i_1}} - \frac{1}{R_{i_j}} \right) - \frac{\beta + \gamma}{R_{i_j}}
\right\}  \\
&= \sum_{j=2}^{n_S} \frac{\alpha_{i_j}}{R_{i_j}} \left\{   \frac{R_{i_j}}{R_{i_1}} - \left(1 + \frac{\beta + \gamma}{\alpha_{i_j}} \right)
\right\}  \\
&\geq 0,
\end{align}
since \eqref{EQ:singleton:Dc:stability:sufficient:condition} holds. Thus, $\sum_{i=1}^n v(S \cap P_i) \geq v(S)$, and hence \eqref{eq:4} is satisfied. The result follows. 
\end{IEEEproof}

\section{Numerical Results}
\label{SC:numerical:results}

We now present numerical results for a network with the default parameter values provided in Table \ref{tab:real_world_params}. We plot the sum of utilities of all users versus different parameters for the following three scenarios: (i) grand coalition, where all users cooperate to form a single coalition, (ii) partition, where users are divided into multiple small coalitions, and (iii) individual downloads, where each user separately downloads the file from the transmitter. 
Unless otherwise stated, we consider $N=20$ users, with the rates $R_1, \ldots, R_N$ given by $20$, $25$, $30$, $35$, $40$, $100$, $105$, $110$, $115$, $120$, $150$, $155$, $160$, $165$, $170$, $200$, $205$, $210$, $215$, $220$, respectively. Also, for scenario (ii) (partition), we adopt the following partition:
\[
\mathbf{P} = \{ \{1,\dots,5\}, \{6,\dots,10\}, \{11,\dots,15\}, \{16,\dots,20\} \},
\]
i.e., the 20 users are divided into four equal-sized coalitions of 5 members each.

\begin{table}[htbp]
\caption{The table shows the default  parameter values used for our numerical results. Each user's valuation, $U_i$, is independently drawn from a uniform distribution with the range $90-100$. Similarly, each user's receive power, $P_{Rx,i}$, is independently drawn from a uniform distribution with the range [0.2W, 0.4W].}
\centering
\begin{tabular}{|c|c|}
\hline
\textbf{Parameter} & \textbf{Value} \\
\hline
Valuation ($U_i$) & 90-100 (variable) \\
\hline
Transmit Power ($P_{Tx}$) & $2$W \\
\hline
Receive Power ($P_{Rx,i}$) & 0.2W-0.4W (variable)  \\
                           
\hline
$a$ & 5 \\
\hline
$b$ & 1.5 \\
\hline
$w$ & 0.5 \\
\hline
File Size ($X$) & 10 \\
\hline
Number of Users ($N$)  & 20 \\
\hline
\end{tabular}
\label{tab:real_world_params}
\end{table}

\begin{figure}
\centering
{\includegraphics[width=0.7\columnwidth]
{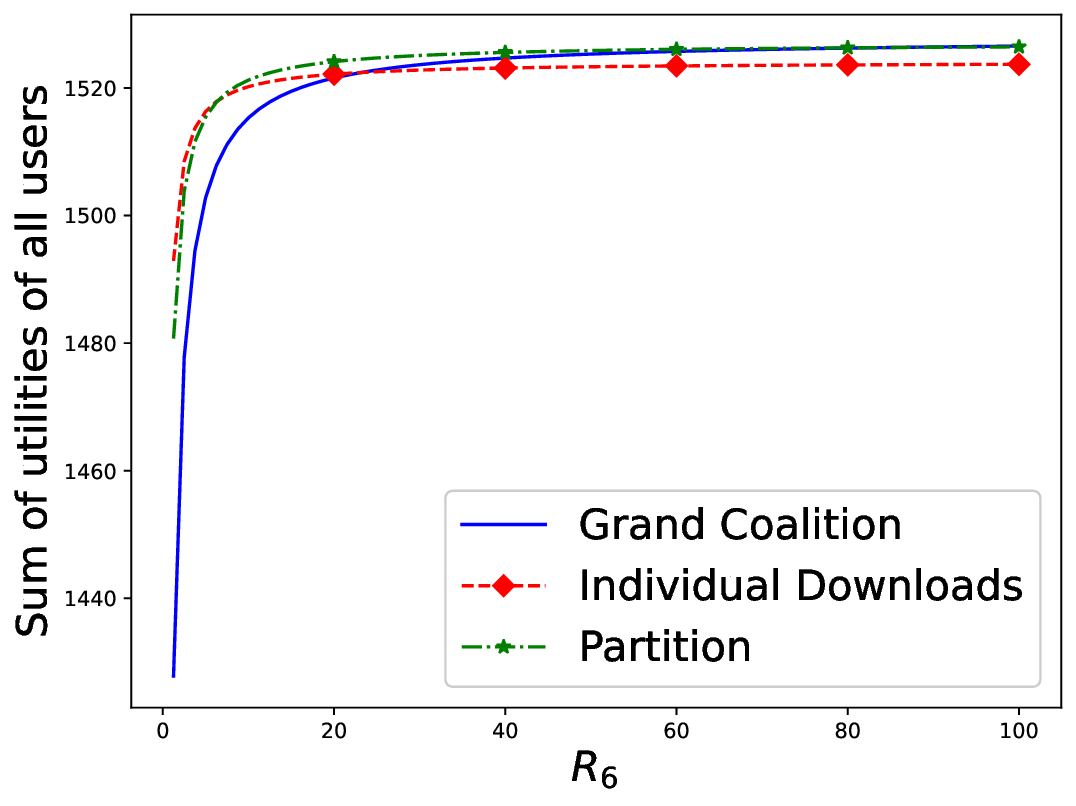}}
     \caption{The figure shows the sum of utilities of all users versus the rate, $R_6$, of the minimum rate user for a scenario in which there are $N=15$ users with rates $R_6, R_7, \ldots,R_{20}$.
      The rates $R_7, \ldots, R_{20}$ remain fixed at their respective values specified in the first paragraph of Section \ref{SC:numerical:results}.}
     \label{fig:1}
    \end{figure}

Fig.~\ref{fig:1} illustrates the variation of the sums of utilities versus the rate, $R_6$, of the minimum rate user. The sums of utilities for all three scenarios increase in  $R_6$ because, by \eqref{EQ:value:function:simplified}, the energy consumed in transmitting and receiving the file and the bandwidth cost decrease. For very low values of $R_6$, the sum of utilities is highest for the individual downloads scenario because under the other two scenarios (partition and grand coalition), due to cooperation with the minimum rate user, users other than the minimum rate user are forced to receive the file at a much lower rate than when they individually download it from the transmitter. Conversely, for very high values of $R_6$, the sum of utilities is highest for the grand coalition because in this case, the reduction in the rate at which users receive the file when they cooperate with the minimum rate user is small, and cooperation is beneficial since it results in savings in the transmission and bandwidth costs. For intermediate values of $R_6$, the sum of utilities is highest for the partition scenario since in this regime, cooperation is moderately beneficial.


\begin{figure}
\centering
    \includegraphics[width=0.7\columnwidth]{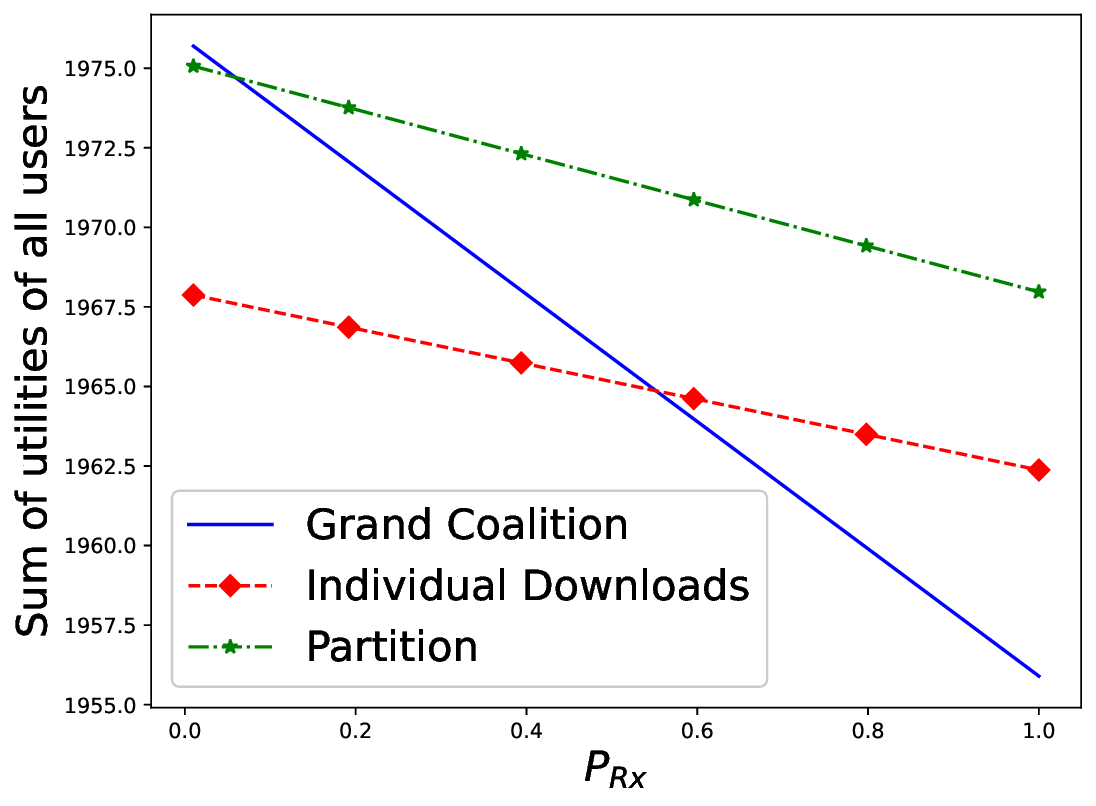}
\caption{The figure shows the sum of utilities of all users versus the receive power, $P_{Rx}$. Here, we have considered a scenario in which  $P_{Rx,i} = P_{Rx}$ for all $i$. }
\label{fig:2}
\end{figure}

Fig.~\ref{fig:2} shows the  variation of the sums of utilities versus the receive power, $P_{Rx}$. The sums of utilities for all three scenarios decrease in  $P_{Rx}$ because, by \eqref{EQ:value:function}, the energy consumed in receiving the file increases. This decrease is the fastest for the grand coalition because the download rates of users in this scenario are the lowest due to cooperation with the minimum rate user, and hence, the energy consumed in receiving the file (see the second term in the RHS of \eqref{EQ:value:function}) increases rapidly in $P_{Rx}$.   

\begin{figure}
\centering
    \includegraphics[width=0.7\columnwidth]{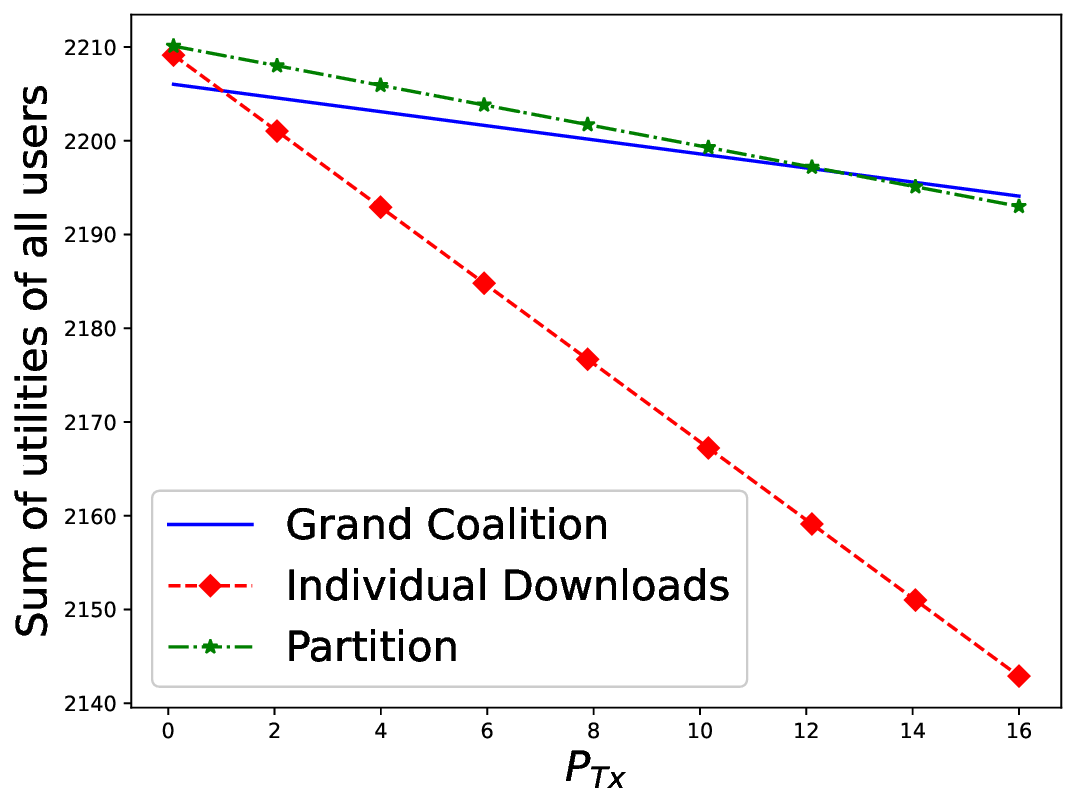}
\caption{The figure shows the sum of utilities of all users versus the transmit power, $P_{Tx}$.}
\label{fig:3}
\end{figure}

Fig.~\ref{fig:3} shows the sums of utilities versus the transmit power, $P_{Tx}$.  As the transmit power increases, the sum of utilities under individual downloads rapidly decreases because in this scenario, a large number of transmissions of the file are required. For sufficiently high values of $P_{Tx}$, the sum of utilities is the highest for the grand coalition because in this scenario, only one transmission of the file is required.

\begin{figure}
\centering
    \includegraphics[width=0.7\columnwidth]{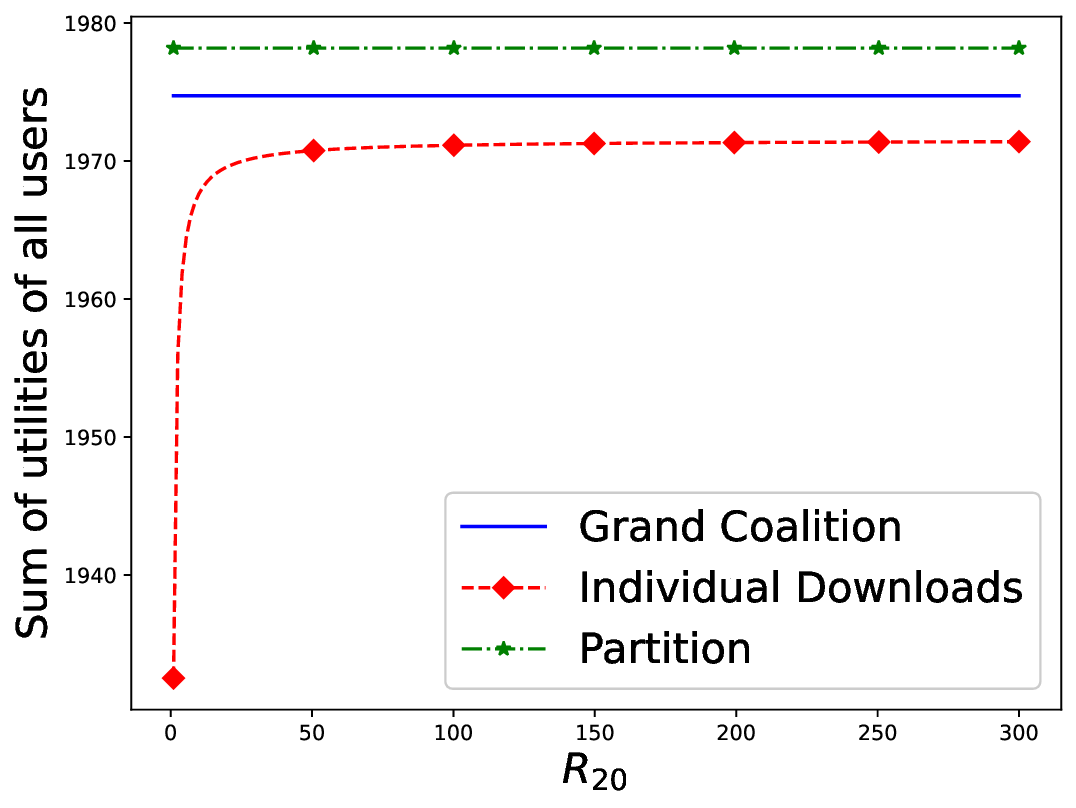}
\caption{The figure shows the sum of utilities of all users versus the rate $R_{20}$.}
\label{fig:4}
\end{figure}

Fig.~\ref{fig:4} shows the sums of utilities versus the rate $R_{20}$. The sum of utilities under individual downloads increases rapidly as $R_{20}$ increases up to about $50$ because, by \eqref{EQ:value:function:simplified}, the energy consumed in transmission and reception of the file to/ at user $20$ and the bandwidth cost decrease. However, as $R_{20}$ increases beyond $50$, this curve becomes nearly flat. This occurs because the cost terms that depend on $R_{20}$ (second and third terms in the RHS of \eqref{EQ:value:function:simplified}) become negligibly small once $R_{20}$ exceeds $50$, due to which further increases in $R_{20}$ have a negligible impact on the sum of utilities under individual downloads. In contrast, the curves for grand coalition and partition  remain almost unchanged as $R_{20}$ increases. This is because in these two scenarios, the file is transmitted to user $20$ not at rate $R_{20}$, but at a lower rate due to the cooperation of user $20$ with other users.

\begin{figure}
\centering
    \includegraphics[width=0.7\columnwidth]{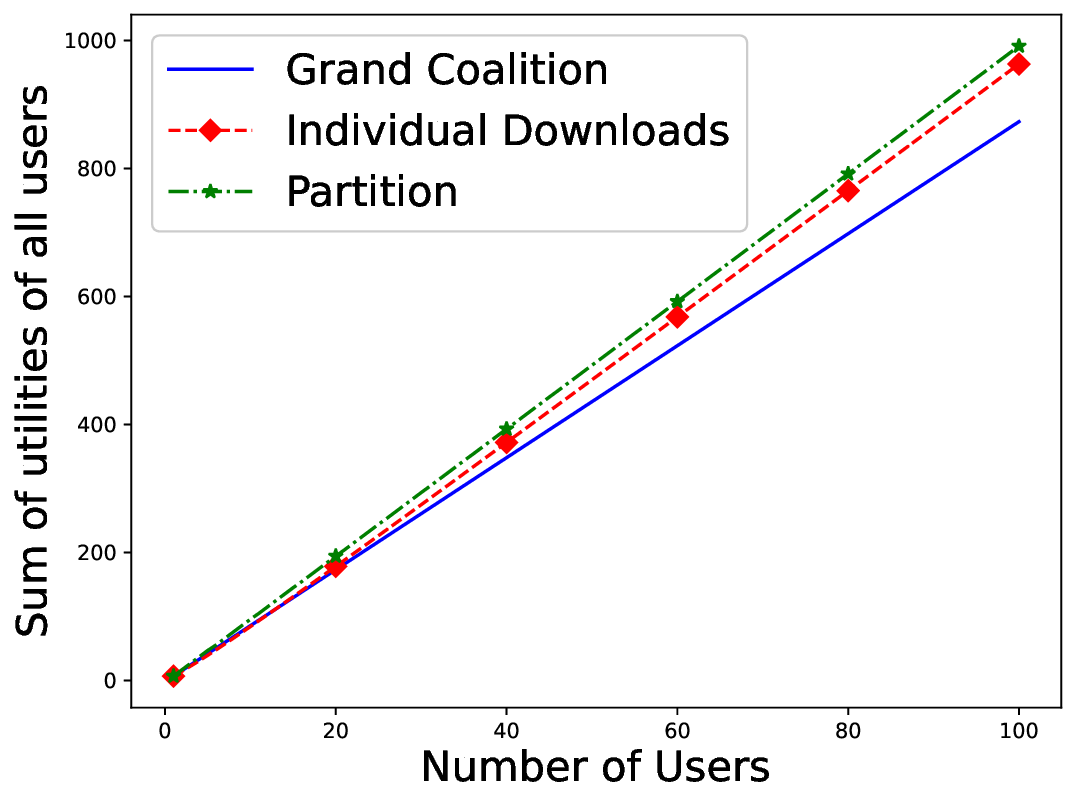}
\caption{The figure shows the sum of utilities of all users versus the number of users, $N$. Here, we have considered a scenario in which  $P_{Rx,i} = P_{Rx}$ for all $i$; also, we have used the value $P_{Rx} = 0.5 \, \mathrm{W}$. The data rates of the users are given by $R_i = B_{\left\lfloor \frac{i}{5} \right\rfloor} + 5\bigl((i-1) \bmod 5\bigr), \qquad i \in \{ 1,\dots,N\},$ where $B_1, B_2, B_3$, and $B_4$ are $20,\;100,\;150$, and $200$, respectively. Note that for $N=20$, these data rates reduce to the default data rates specified in the first paragraph of this section. For the ``Partition" curve, users are partitioned sequentially into disjoint coalitions of five users each, i.e., $\{1,\dots,5\}, \{6,\dots,10\}, \dots$.}
\label{fig:5}
\end{figure}

Fig.~\ref{fig:5} shows the sums of utilities versus the number of users, $N$. 
As $N$ increases, the sums of utilities in all three scenarios increase because the increase in the sum of valuations (first term in the RHS of \eqref{EQ:value:function:simplified}) is greater than the increase in the energy costs of transmission and reception and the bandwidth costs.

\begin{figure}
\centering
    \includegraphics[width=0.7\columnwidth]{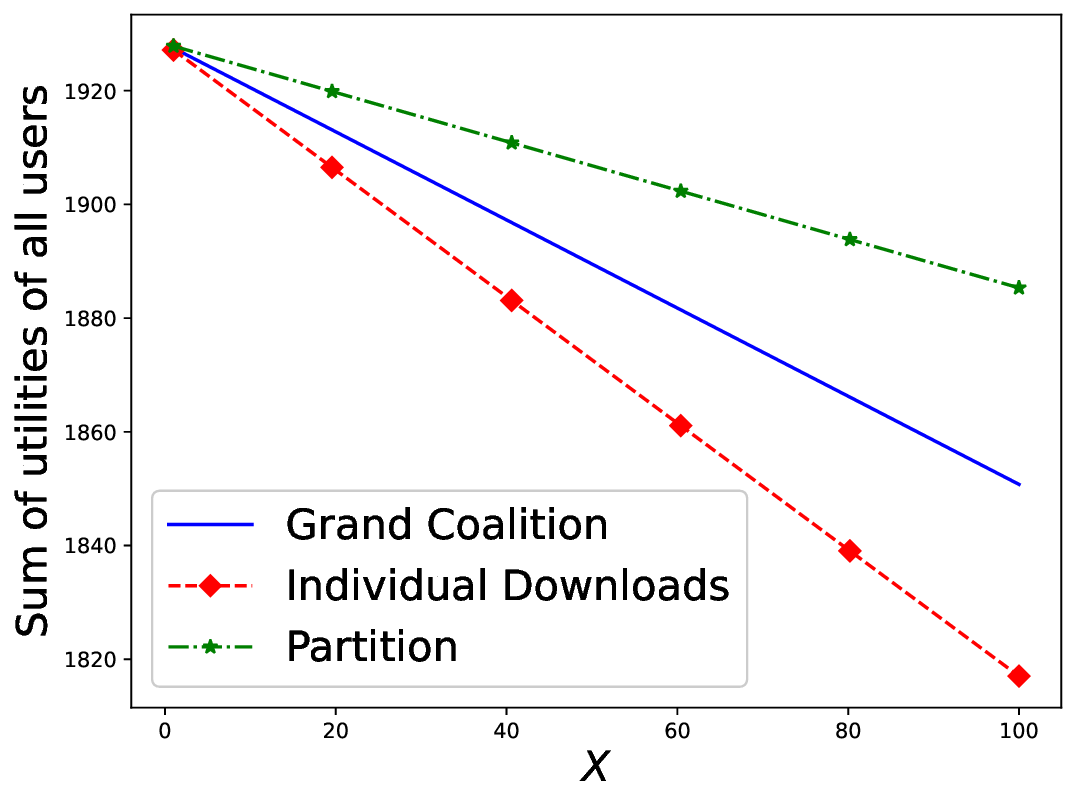}
\caption{The figure shows the sum of utilities of all users versus the file size, $X$.}
\label{fig:6}
\end{figure}

Finally, Fig.~\ref{fig:6} shows the sums of utilities versus the file size $X$. As $X$ increases, the sums of utilities in all three scenarios linearly decrease because the  transmission and reception energy costs and the bandwidth costs (see the second, third, and fourth terms in the RHS of \eqref{EQ:value:function}) are directly proportional to $X$. When the file size $X$ is large, individual downloads become highly inefficient because the large number of transmissions require a significant amount of energy and bandwidth, making this the least favorable option. Under the grand coalition, the number of transmissions is reduced to one, but its efficiency is limited by the fact that transmission takes place at a very low data rate. The partition approach balances both aspects: the required number of transmissions is lower than in the individual downloads scenario and transmission of the file takes place at higher rates than in the grand coalition scenario.

\section{Conclusions and Future Work}
\label{SC:conclusions:future:work}
Using the framework of cooperative game theory, we investigated conditions under which users have incentives to cooperate among themselves to form coalitions for the purpose of receiving a popular file via multicast from a transmitter.  First, using the solution concept of core, we investigated conditions under which it is beneficial for all users to cooperate, i.e., the grand coalition is stable. We provided several sets of sufficient conditions under which the core is non-empty as well as those under which the core is empty. Next, we used the concept of $\mathbb{D}_c$-stability to identify a set of sufficient conditions under which the users in the network form a certain fixed number of coalitions such that all the users within each coalition cooperate among themselves. We also studied cooperation among different users using numerical computations. The problem of coalition formation in the context of multicast addressed in this paper is fundamental, and our analysis provides new insights into the feasibility of stable cooperative multicast strategies, contributing to a deeper understanding of cooperation in wireless networks. A direction for future research is to investigate, using solution concepts such as the Shapley value and nucleolus, different ways in which the value of a coalition formed for multicast can be distributed among the users of the coalition in an efficient and fair manner.



\bibliographystyle{IEEEtran}

\bibliography{ref}
\nocite{*}

\end{document}